%% file: main.tex
\documentclass[a4paper,11pt]{article}
\usepackage{pos}
\input{shortcuts}

\title{Approaches to the Inverse Problem}
\ShortTitle{Approaches to the Inverse Problem}

\author[a]{Luigi Del Debbio}
\author*[b]{Alessandro Lupo}
\author[c]{Marco Panero}
\author[d]{Nazario Tantalo}

\affiliation[a]{Higgs Centre for Theoretical Physics, School of Physics \& Astronomy, The University of Edinburgh, Peter Guthrie Tait Road, Edinburgh EH9 3FD, United Kingdom}

\affiliation[b]{Aix Marseille Univ, Université de Toulon, CNRS, CPT, Marseille, France}

\affiliation[c]{Department of Physics, University of Turin \& INFN, Turin,
Via Pietro Giuria 1, I-20125 Turin, Italy}

\affiliation[d]{University and INFN of Roma Tor Vergata,
	Via della Ricerca Scientifica 1, I-00133, Rome, Italy}

\emailAdd{alessandro.lupo@cpt.univ-mrs.fr}

\abstract{
The analytical continuation of correlation functions from imaginary to real time is a crucial step in lattice gauge theories, and it challenges our ability to derive non-perturbative predictions from lattice simulations. We review aspects of this ``inverse problem'', which has driven both theoretical and algorithmic advancements in recent years, opening new promising avenues for non-perturbative studies of the Standard Model and beyond. The focus of this proceeding will be on variations of Backus-Gilbert and Bayesian methods.
}

\FullConference{European network for Particle physics, Lattice field theory and Extreme computing (EuroPLEx2023)\\
 11-15 September 2023\\
Berlin, Germany\\}

\begin{document}
\maketitle

\section{Introduction}
Lattice methods provide a systematically improvable framework to study gauge theories non-perturbatively, and from first principles. A vast number of increasingly precise results have been obtained for Quantum Chromodynamics (QCD) and for strongly-interacting gauge theories in general. Regarding the time coordinate as a complex variable, lattice simulations allow computing functions of time on the imaginary (Euclidean) axis. The analytical continuation back to real (Minkowski) time is close to the problem of solving a Fredholm integral equation, with a number of additional difficulties originating from the nature of lattice data. The problem is in fact, in this context, ill-defined and ill-conditioned. Nonetheless, given its importance, a large effort has been devoted to developing regularising strategies~\cite{Asakawa:2000tr, Burnier:2013nla, Hansen:2019idp, Kades:2019wtd, Horak:2021syv, Bailas:2020qmv, Bergamaschi:2023xzx, Buzzicotti:2023qdv, Bruno:2024fqc, DelDebbio:2024lwm} to solve what is generally referred to as the ``inverse problem''. The motivations are manifold: a clean solution with quantifiable errors would allow ab-initio calculations of phenomenologically important quantities such as scattering amplitudes~\cite{Barata:1990rn, Bulava:2019kbi, Patella:2024cto}, inclusive decay rates~\cite{Gambino:2020crt, Bulava:2021fre, Gambino:2022dvu,  ExtendedTwistedMassCollaborationETMC:2022sta, Evangelista:2023fmt, Barone:2023tbl, Alexandrou:2024gpl}, as well as predictions from QCD at a finite temperature~\cite{Meyer:2007ic, Meyer:2007dy, Aarts:2007wj, Caron-Huot:2009ncn, Meyer:2011gj, Aarts:2011sm, Aarts:2012ka, Aarts:2013kaa, Aarts:2014nba, Ding:2015ona, Rothkopf:2019ipj, Itou:2020azb, Altenkort:2022yhb, Bonanno:2023ljc, Bonanno:2023thi, Aarts:2023vsf}.

In the following, after a brief introduction to the problem, we shall review two closely related methods that have been recently devised in order to provide a solution to the inverse problem with controlled systematics. The HLT procedure~\cite{Hansen:2019idp}, close to Backus-Gilbert methods, and its Bayesian formulation given in Ref.~\cite{DelDebbio:2024lwm}, which differs from other approaches previously devised based on Bayesian inference. Moreover, we will show a method to systematically validate a given approach to the inverse problem given a set of correlators for which we can measure the covariance matrix.

In the context of lattice gauge theories, the inverse problem refers to the operation of extracting the spectral density from a correlation function of gauge-invariant operators. In the case of a function of two operators separated by the Euclidean time $t$, the relation can be written as
\begin{equation}\label{eq:introduce_inverse_problem}
    C_{LT}(t) = \int_{0}^\infty dE \; b_T(t,E) \, \rho_{LT}(E) \; ,
\end{equation}
where $C_{TL}$ and $\rho_{TL}$ are respectively the correlator and the spectral density in a four dimensional lattice, with spatial volume $L^3$, time extent $T$, and lattice spacing $a$. The function $b_T(t,E)$ can be written, in general, as
\begin{equation}
    b_T(t,E) = b_+ e^{-tE} + b_- e^{-(T-t)E} \, ,
\end{equation}
the numbers $b_{\pm}$ are integer numbers that depend on the nature of the correlation function.  The distributional nature of the spectral function requires the introduction of a Schwartz function $S_\sigma(E-\omega)$ and a corresponding smeared spectral function:
\begin{equation}\label{eq:smearing}
    \rho_L(\sigma;\omega) = \int dE \, \mathcal{S}_\sigma(\omega-E) \, \rho_L(E) \, .
\end{equation}
The parameter $\sigma$ allows controlling the size of the smearing. For practical use, the Schwartz function should be defined in such a way that $\lim_{\sigma \rightarrow 0} \mathcal{S}_\sigma(\omega-E) = \delta(\omega-E)$. In the following, we shall introduce so-called linear methods, where the smeared version of the solution is represented as a linear combination of correlation functions:
\begin{equation}\label{eq:linear_problem}
    \rho_L(\sigma, \omega) = \sum_{\tau = 1}^{\tiny \tau_{\rm max} } g_t(E) C_L(a \tau) \, , \;\;\;\;\; 0 < \tau \leq \tau_{\max} \; , \;\;\; \;\; \tau = t / a \; .
\end{equation}
This class of solutions always provides a smeared spectral function. From Eqs.~\eqref{eq:linear_problem} and~\eqref{eq:smearing} it in fact follows that
\begin{equation}\label{eq:smearing_kernel_def}
    \mathcal{S}_\sigma(E,\omega) = \sum_{\tau =1}^{\tau_{\rm max}} g_\tau(\sigma;\omega) \,  b_T(a \tau,E)  \; .
\end{equation}

A number of complications arise when dealing with the inverse problem. While  the spectral function has in principle support over the real axis, the available data constitutes a discrete and finite set, in an apparent mismatch of information. The smearing operation, however, already eases this aspect, and the finiteness of the dataset only amounts to a systematic error, without compromising the uniqueness of the solution. Another complication, typical to lattice data, is the exponential dumping of the signal due to the functions $b_T(t,E)$, which reduces to $\exp(-tE)$ at large $T$. As a consequence, linear solutions to the inverse problem give rise to ill-conditioned systems which require very precise input data. Moreover, the unavoidable presence of errors on the input correlators results in numerical instabilities of the solution, possibly posing the most serious challenge. On the lattice, this is additionally exacerbated by the ill conditioning of the system. The regularisation that is needed to stabilise the problem introduces a bias that is hard to estimate. In discussing approaches to the inverse problem we will focus, in the following, on how they manage the aforementioned issues. After introducing the procedures and the connection between them, we shall comment on their ability to provide unbiased solutions.

\section{The HLT procedure}
\label{sec:hlt}

We briefly review this procedure which provides a set of rules for the calculation of smeared spectral functions and the treatment of its errors, see ~\cite{Hansen:2019idp, DelDebbio:2024lwm} for more comprehensive discussions. The mathematics of the procedure shares similarities with other methods that can be found in particle physics~\cite{FURMANSKI1982237}, as well as in geophysics~\cite{10.1111/j.1365-246X.1968.tb00216.x, pijpers1994sola, zaroli2016}. 

The goal is to find the best approximation for a spectral function smeared with a chosen kernel $\mathcal{S}_\sigma(E,\omega)$. To this end, let us introduce a representation of the smearing kernel based on the functions $b_T(t,E)$:
\begin{equation}\label{eq:exact_kernel_representation}
    \mathcal{S}_\sigma(E-\omega) = \sum_{\tau = 1}^{\infty} \hat{g}_\tau(\sigma; \omega) b_\infty(a \tau, E) \, .
\end{equation}
Let $C(a \tau)$ be the exact correlators (without errors). The spectral function smeared with $\mathcal{S}_\sigma$ is then
\begin{equation}\label{eq:exact_rhos_representation}
    \rho(\sigma; \omega) = \sum_{\tau = 1}^{\infty} \hat{g}_\tau(\sigma; \omega) C(a \tau) \, .
\end{equation}
Both Eqs.~\eqref{eq:exact_kernel_representation} and ~\eqref{eq:exact_rhos_representation} are exact representations, and an expression for the coefficients $\hat{\vec{g}}$ can be given:
\begin{equation}
    \hat{\vec{g}}(\sigma;\omega) = \argmin_{\vec{g} \in \mathbb{R}^{\mathbb{N}}} \int_0^\infty dE \, e^{\alpha E} \left| \sum_{\tau=1}^{\infty} g_\tau(\omega) b_\infty(a \tau, E) - \mathcal{S}_\sigma(E,\omega) \right|^2 \, ,
\end{equation}
where each real value of $\alpha<2$ defines norms with different weights. In practice, one has access to a finite number $\tau_{\rm max}$ of correlators. An approximation for the coefficients, that we label $\vec{g}$,  can still be given with a similar expression:
\begin{equation}\label{eq:coeff_approx_A_only}
        \vec{g}(\sigma;\omega) = \argmin_{\vec{g} \in \mathbb{R}^{\tau_{\rm max}}} \int_0^\infty dE \, e^{\alpha E} \left| \sum_{\tau=1}^{\tau_{\rm max}} g_\tau(\omega) b_T(a \tau, E) - \mathcal{S}_\sigma(E,\omega) \right|^2 \, .
\end{equation}
The approximated kernel and smeared density are:
\begin{equation}\label{eq:approx_kernel_sol}
    \mathcal{S}_\sigma(E-\omega) \simeq \sum_{\tau = 1}^{\tau_{\rm max}} g_\tau(\sigma; \omega) b_\infty(a \tau, E) \, , \;\;\;\;\;\;\;\;\;\;\; \rho(\sigma; \omega) \simeq \sum_{\tau = 1}^{\tau_{\rm max}} g_\tau(\sigma; \omega) C(a \tau) \, .
\end{equation}
The representations in Eq.~\eqref{eq:approx_kernel_sol} are in fact still valid up to a systematic error, which was shown to vanish quickly, being small even for moderate values of $\tau_{\rm max}$, see e.g. Ref.~\cite{Hansen:2019idp}. Extracting a continuous function from a finite number of data does not seem to make our problem ill-defined, as it merely amounts to a systematic error. The reason is, however, that the problem is already regulated: we are not accessing the underlying spectral function but a smeared version of it.

In realistic cases, the correlator is affected by many sources of error, statistical and systematic. This makes the problem ill-posed even in presence of the smearing, because the representations of Eq.~\eqref{eq:approx_kernel_sol} are not stable within error fluctuations of the correlation function. An additional regulator needs to be applied. Following the prescription of Backus and Gilbert, we introduce the regulated coefficients, that we keep calling $\vec{g}$ with abuse of notation:
\begin{multline}\label{eq:regulated_coefficients}
        \vec{g}(\sigma;\omega) = \argmin_{\vec{g} \in \mathbb{R}^{\tau_{\rm max}}} \int_0^\infty dE \, e^{\alpha E} \left| \sum_{r=1}^{\tau_{\rm max}} g_\tau(\omega) b_T(a \tau, E) - \mathcal{S}_\sigma(E,\omega) \right|^2 \\  + \lambda \; \sum_{\tau_1, \tau_2=1}^{\tau_{\rm max}} g_{\tau_1}(\omega) \, B_{\tau_1 \tau_2}\,  g_{\tau_2}(\omega) \, , \;\;\;\;\; \lambda \in (0,\infty)
\end{multline}
The matrix $B$ is the covariance matrix of the correlator, often rescaled to be dimensionless. The parameter $\lambda$ regulates the coefficients by making them smaller, which stabilises Eq.~\eqref{eq:approx_kernel_sol}. This introduces a bias, since the coefficients will now differ from the approximation of Eq.~\eqref{eq:coeff_approx_A_only}. In the limit $\lambda \rightarrow \infty$ the bias is maximum, and the solution is simply $\vec{g} = \vec{0}$, while at $\lambda=0$ the bias is not present, but the statistical noise is, effectively, infinitely large.

In order to make the prediction for the smeared spectral density reliable, an intermediate value of $\lambda$ must be chosen, and the effect of its resulting bias needs to be assessed precisely. The procedure consists in performing a scan of different values of $\lambda$, starting with a larger value and gradually decreasing it until the output is dominated by noise. If the signal is good enough, one can identify a plateau at smaller values of $\lambda$, before the signal is lost into the statistical noise. This strategy has analogies with fits in Euclidean time of correlation functions and effective masses. Numerical work on both synthetic and real Monte Carlo data has been supporting the capability of this procedure to provide unbiased results, with a notable example being Ref.~\cite{Bulava:2021fre}, where lattice data was tested against analytic results.

\section{Bayesian methods: interplay between smearing and priors}

The inverse problem has been attacked with Bayesian frameworks for many years~\cite{Asakawa:2000tr, Burnier:2013nla}. It is a powerful tool with many perks, such as the ability to incorporate prior knowledge about the solution, and to produce analytic expressions for the results, which can add precious insights. Here, we shall discuss the setup based on Gaussian Processes (GP). We will focus on the interplay between the choice of the prior and the smearing, and the treatment of the bias introduced while regularising the problem, in order to compare with Section~\ref{sec:hlt}.

There are at least two ways of setting up the Bayesian solution to the inverse problem. The common ground, and a difference with respect to frequentist methods (e.g. Backus-Gilbert), is that one computes probability densities associated with the spectral functions, rather than the spectral functions themselves. A first strategy, which has been very popular, is to compute the probability density for the spectral function directly. In this case, the smearing arises implicitly because we are numerically forced to give a width to the covariance $\mathcal{K}^{\rm prior}$. Alternatively, one we can set up the problem to compute the probability density for a spectral function that is smeared with a chosen kernel, similarly to what is done with the HLT method. In the latter case, the smearing is explicit, and we are able to make consistent assumptions about the functional nature of the prior and posterior distributions. We shall now discuss both cases.

\paragraph{Implicit smearing.} Leaving out the mathematical details, that can be found in our recent work~\cite{DelDebbio:2024lwm}, we briefly mention that the spectral density is promoted to a stochastic field $\mathcal{R}(E)$, that we assume to have a Gaussian underlying distribution $\Pi[\mathcal{R}]$. The latter is fully specified by a central value and a covariance for the field variable:
\begin{equation}
    \rho^{\rm prior}(\omega) = \int \mathcal{D}\mathcal{R}\, \Pi[\mathcal{R}] \; \mathcal{R}(\omega)  \; .
\end{equation}
\begin{equation}
    \mathcal{K}^{\rm prior}(\omega_1, \omega_2) = \int \mathcal{D}\mathcal{R}\, \Pi[\mathcal{R}] \; \mathcal{R}(\omega_1) \mathcal{R}(\omega_2)  \; .
\end{equation}
The observational noise is also assumed to be a random vector that is being sampled from a multivariate Gaussian distribution: in this case the mean is assumed to be zero, and the covariance matrix $\text{Cov}_d$ is measured from the data. By using Bayes' theorem, we can then obtain a posterior distribution that corresponds to the prior one, conditioned by the data, see Ref.~\cite{DelDebbio:2024lwm} for details. Since every distribution is Gaussian, the posterior will be as well. The central value is the function
\begin{equation}\label{eq:naive_gp_mean}
    \rho^{\rm post}(\omega) = \sum_{\tau = 1}^{\tau_{\rm max}} g^{\rm GP}_{\tau}(\omega) C(a \tau) \; , 
\end{equation}
and the variance is
\begin{equation}\label{eq:naive_gp_variance}
    \mathcal{K}^{\rm post}(\omega,\omega) =
     \mathcal{K}^{\rm prior}(\omega,\omega) - \sum_{\tau = 1}^{\tau_{\rm max}} g_\tau^{\rm GP}(\omega)  \int_0^\infty dE \, \mathcal{K}^{\rm prior}(\omega,E) b_T(a \tau, E) \; .
\end{equation}
In the above equations we have made the common choice to set $\rho^{\rm prior}=0$. The coefficients $g^{\rm GP}_\tau$ are given in Table~\ref{tab:details}. The smearing kernel is not known a priori, it is defined implicitly by Eq.~\eqref{eq:approx_kernel_sol} and it can be computed once the coefficients are known. However, one can a priori acquire a qualitative understanding of how the smearing kernel may look, based on the shape of the prior. The prior covariance of the spectral function, in fact, determines certain properties of the posterior, such as its typical correlation length. In a finite volume, the spectral density is a sum of $\delta$-functions. The central value of the posterior distribution \textit{should accordingly be a distribution}, which can only happen in the limiting case in which the prior itself is infinitely narrow. This singular behaviour cannot be achieved numerically, and one settles for a prior with a finite width. As a consequence, the spectral density is smeared, and the resulting kernel has a typical width, that is loosely determined by the width of the prior. Interestingly, in the limit in which $\mathcal{K}^{\rm prior}$ becomes a $\delta$-function, the coefficients are identical to those defined in Eq.~\eqref{eq:regulated_coefficients} at $\alpha=0$ and in the limit of vanishing $\sigma$, see Table~\ref{tab:details}.

\paragraph{Explicit smearing.} We consider the problem of determining the probability density of the field variable $\mathcal{R}_\sigma$ describing a spectral density smeared with a given Schwartz function $S_\sigma$:
\begin{equation}
    \mathcal{R}_\sigma(\omega) = \int_0^\infty dE \, \mathcal{S}_\sigma(\omega-E) \mathcal{R}(E) \, .
\end{equation}
Applying Bayes' theorem as before, the posterior probability density for the spectral density smeared with $\mathcal{S_\sigma}$ can be computed. The resulting expression for the coefficients can be found in Table~\ref{tab:details}, and more details are in Ref.~\cite{DelDebbio:2024lwm}.

As stated above, the finite-volume spectral density is characterised by sharp peaks, and the choice of the covariance of the posterior should allow for this behaviour. A natural choice is then to use a fully uncorrelated covariance,
\begin{equation}\label{eq:the_covariance_is_a_delta}
    \mathcal{K}^{\rm prior}(E,\omega) = \frac{ e^{\alpha E}}{\lambda} \, \delta(E-\omega) \, .
\end{equation}
This can now be done, because the spectral density appears at every stage in a convolution with the Schwartz function $S_\sigma$, and no singular behaviour is encountered at finite $\sigma$. The choice of the normalisation of the covariance allows for a direct connection with the HLT procedure. In fact, using Eq.~\eqref{eq:the_covariance_is_a_delta} one obtains a posterior distribution for $\mathcal{R}_\sigma$ characterised by a mean function that is identical to the HLT solution, i.e. with coefficients given by Eq.~\eqref{eq:regulated_coefficients}. The covariance of the posterior distribution is given by 
\begin{equation}\label{eq:Kpost}
\mathcal{K}^{\rm post}(\omega,\omega) = \int dE \,  \frac{e^{\alpha E}}{\lambda} \, \mathcal{S}_\sigma(\omega,E) 
\left(  \sum_{\tau=1}^{\tau_{\rm max}} g_\tau(\sigma, \omega) \, b_T(a \tau,E) - \mathcal{S}_\sigma(\omega,E) \right) \, ,
\end{equation}
which does not have an analogue in any frequentist approach. The square root of Eq.~\eqref{eq:Kpost} is however interpreted as the statistical error on the smeared spectral density, and in this respect it can be compared to a frequentist error, which can be estimated, for instance, with a resampling procedure.

\begin{table}[htb]
    \centering
    \renewcommand{\arraystretch}{1.2}
    \resizebox{\textwidth}{!}{
\begin{tabular}{|c|c|c|}
    \hline
    \multicolumn{3}{|c|}{$g_{\tau_1}(\omega) = \sum_{\tau_2} (\Sigma + \lambda B)^{-1}_{\tau_1 \tau_2}  \; f_{\tau_2}(\omega)$} \\ \hline
    & $\Sigma_{\tau_1 \tau_2}$ & $f_\tau(\omega)$ \\[4pt] \hline
    {\small HLT} & $\int_{E} b_T(\tau_1,E) b_T(\tau_2,E)$ &  $\int_E \mathcal{S}_\sigma(\omega-E)  b_T(\tau,E)$  \\[1pt] \hline
    {\small GP} {\tiny impl. smr.}& $\int_{E_1 E_2} b_T(\tau_1, E_1)  \mathcal{K}^{\rm prior}(E_1,E_2) \, b_T(\tau_2, E_2) $ & $\int_E \, \mathcal{K}^{\rm prior}(\omega-E)  b_T(\tau,E)$ \\[1pt] \hline
    {\small GP} {\tiny expl. smr.} & $ \int_{E_1 E_2} b_T(\tau_1, E_1)\mathcal{K}^{\rm prior}(E_1,E_2) b_T(\tau_2, E_2) $ & $\int_{E_1 E_2} b_T(\tau, E_1) \, \mathcal{K}^{\rm prior}(E_1, E_2) \, \mathcal{S}_\sigma(E_2,\omega)$\\[1pt] \hline
\end{tabular}
}
    \caption{Formulae for the linear systems determining the coefficients in the three cases analysed in this work: HLT, Bayesian inference with GPs with implicit (induced by $\mathcal{K}^{\rm prior}$) and explicit (with the smearing kernel $\mathcal{S}_\sigma$) smearing of the spectral function. The matrix $B$ is proportional to the covariance of the correlator.}
    \label{tab:details}
\end{table}

\paragraph{Regularising bias.} In either of the Bayesian setups that we described, as far as the stability with respect to the noise on the input data is concerned, the problem is regularised in the same way as it is in Backus-Gilbert methods. The expression for the coefficients $g_\tau$ that is explicitly given in Tab.~\ref{tab:details} involves a linear system, where an ill-conditioned matrix is regularised by adding the covariance of the data, times a parameter. In Backus-Gilbert methods, this is done by hand, as shown in Eq.~\eqref{eq:regulated_coefficients}. With Bayesian methods, it appears as a hyperparameter which normalises the covariance of the prior, as seen in Eq.~\eqref{eq:the_covariance_is_a_delta} (the same holds even if the covariance of the prior is not a $\delta$-function). The value of $\lambda$, and of any other hyperparameter, is selected as the one that maximises a likelihood function which represents the probability of observing the data, given a choice of the hyperparameters. It is natural to ask how this compares to the stability analysis of the HLT method. In order to make a precise comparison, we use the Bayesian setup with explicit smearing: in this case, in fact, the spectral function is smeared with the same kernel (a Gaussian function), and a comparison is straightforward. In Ref.~\cite{Buzzicotti:2023qdv} a comparison between this setup and HLT was done using lattice data. We have found that values of $\lambda$ in the stability region tend to be those that maximise the likelihood, or equivalently minimise its negative logarithm (NLL). This suggests that the two approaches are consistent. In the same work, we compared the size of the error, which is the other notable difference in the two approaches. Our findings suggest that at fixed $\lambda$ the Bayesian error is more conservative.

\section{Systematic validation against pseudodata}

\begin{figure}[t]
    \centering
    \includegraphics[width=\textwidth]{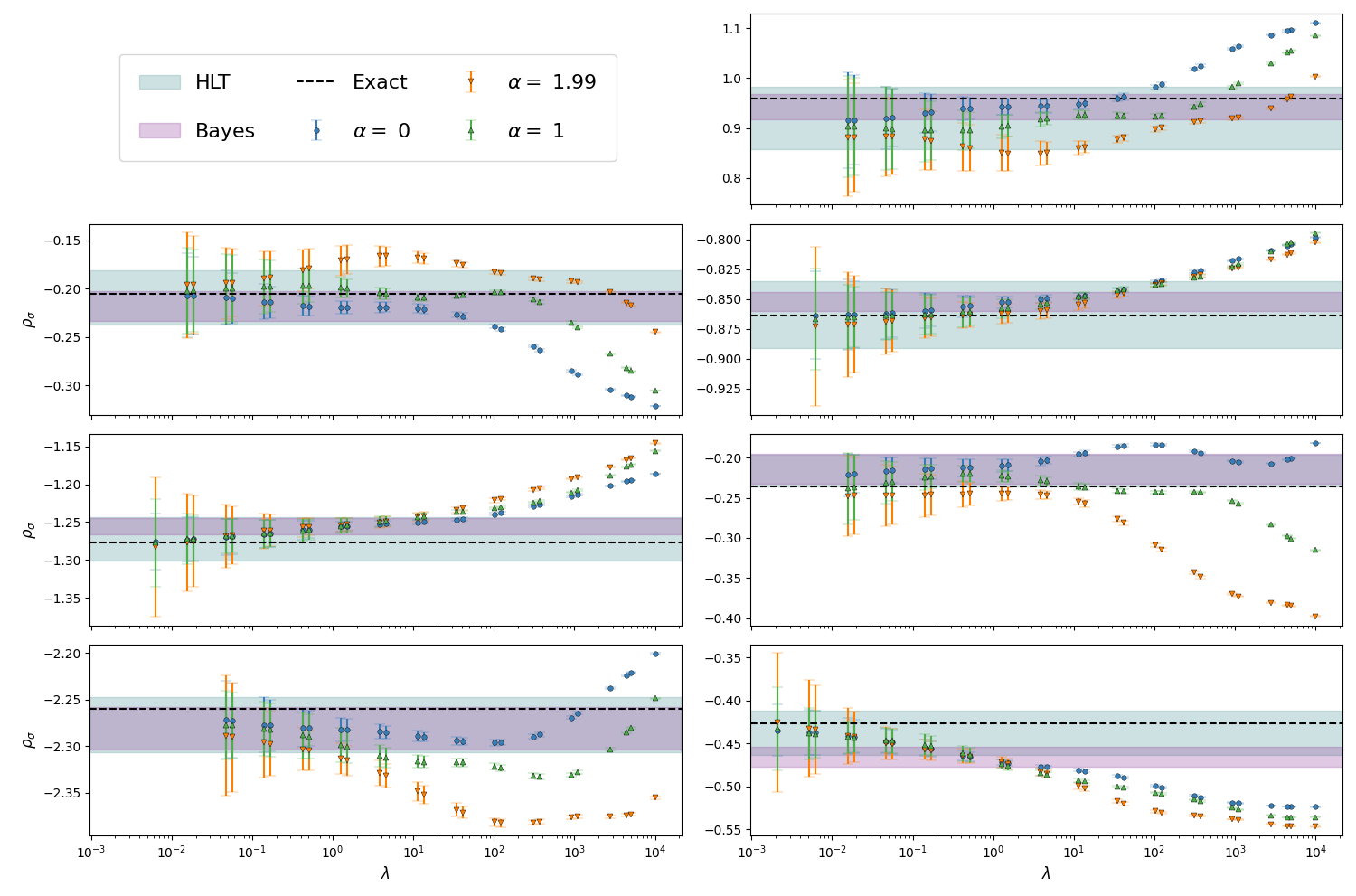}
    \caption{Examples of the stability analysis performed in different instances of the inverse problem. The exact result is also reported as a black line. The HLT band corresponds to the value determined by an automated plateau selection. The Bayes' band is selected by minimising the NLL.}
    \label{fig:lambda_scan_instances}
\end{figure}

With an increasing variety of methods available, it is desirable to devise a flexible validating strategy that can allow making comparisons between different approaches to the inverse problem. At the same time, it should be taken into account that features of lattice data can vary significantly depending on the details of the simulations: the optimal strategy can vary. Moreover, a general question about the possibility itself to provide unbiased estimation of the spectral function remains, since the problem is ill-posed to begin with. The arguments raised with the stability analysis, or the minimisation of the NLL support a positive answer to this question, which can be made more quantitative with a validation strategy.

With these motivations, we defined $N_{\rm toys}$ sets of synthetic finite-volume spectral functions, 
\begin{equation}
     \rho(E) = \sum_{n=0}^{n_{\rm max}-1} w_n \; \delta(E - E_n \;) , \;\;\;\;\; E_0 < E_1 \leq \dots \; 
\end{equation}
which are specified by an equivalent number of sets for the energies $E_n$ and matrix elements $w_n$. The value of $n_{\rm max}$ has been chosen to be roughly the number of finite-volume states populating the energy range of interest for a given volume. We decided for simplicity to take the energy levels to be equally spaced. The matrix elements are instead sampled from a multivariate normal distribution, specified by a vanishing mean, and a covariance given by
\begin{equation}\label{eq:toy_cov}
    K_{\rm weights}(n, n') = \kappa  \;  \exp \left( - \dfrac{(E_n-E_{n'})^2}{2 \epsilon^2} \right) \, .
\end{equation}
The value of $\epsilon$ is set to be smaller than the spacing between subsequent energy levels, in order to reproduce the sharply decorrelated features of a finite-volume spectral function. The parameter $\kappa$ is free, and different values can be tested. For each of the $N_{\rm toys}$ sets of $\{E_n, w_n\}$, the corresponding correlator can be defined as 
\begin{equation}\label{eq:corr_rnd_peaks}
    C(t) = \sum_{n=0}^{n_{\rm max}-1} w_n \; e^{-|t|E_n} \; , \;\;\;\;\; E_0 < E_1 \leq \dots \; .
\end{equation}
Statistical noise is the injected by generating $N_{\rm cnfg}$ configurations for the correlator, using a multivariate normal distribution that is based on the covariance matrix of the correlator that was measured on the lattice. This step is crucial in order to ensure the generated statistics related to the one of the desired lattice simulations. With this setup, one can solve $N_{\rm toys}$ instances of the inverse problem and check the results against known solutions.

\begin{figure}[bh!]
    \centering
    \includegraphics[width=0.49\textwidth]{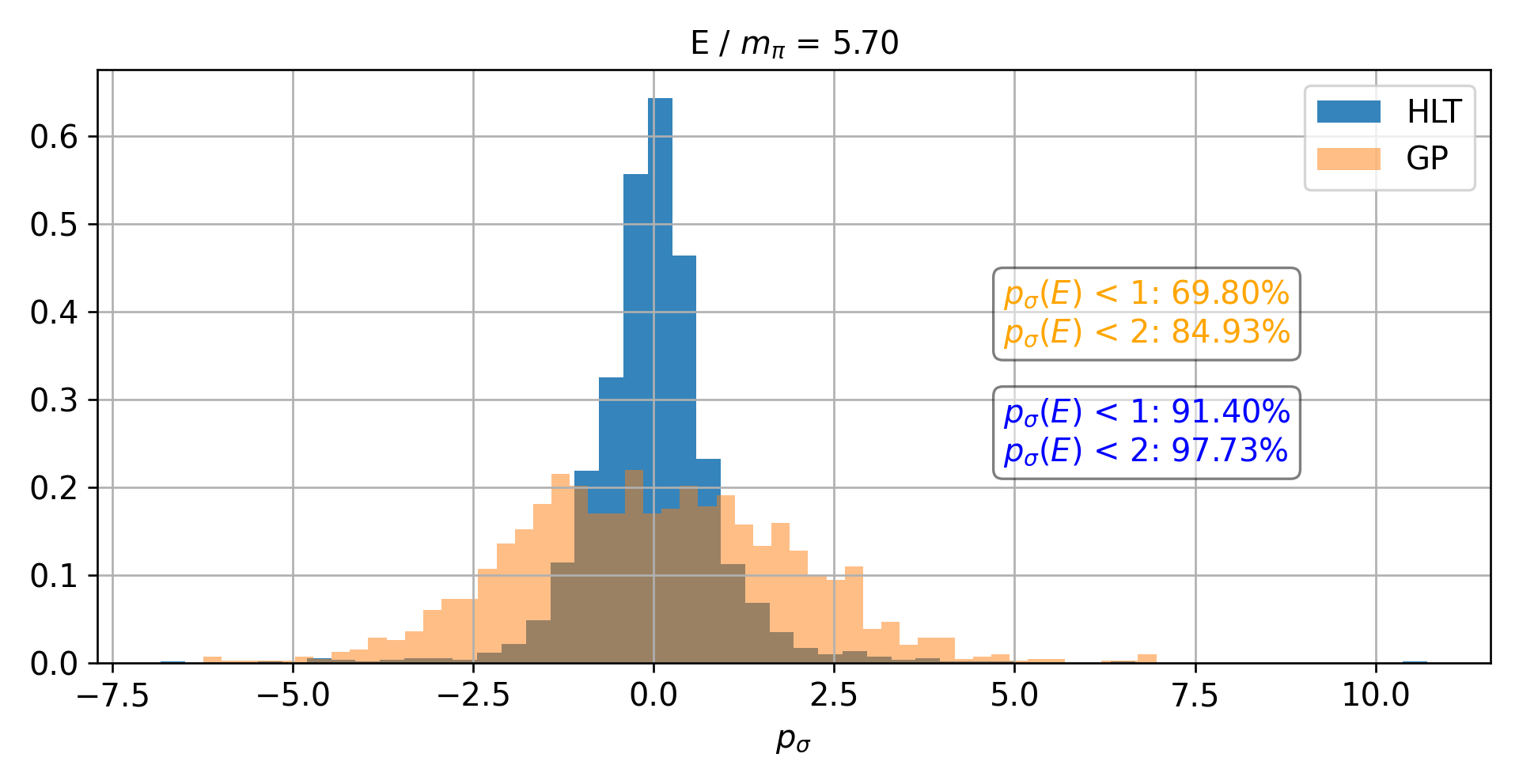}
    \includegraphics[width=0.49\textwidth]{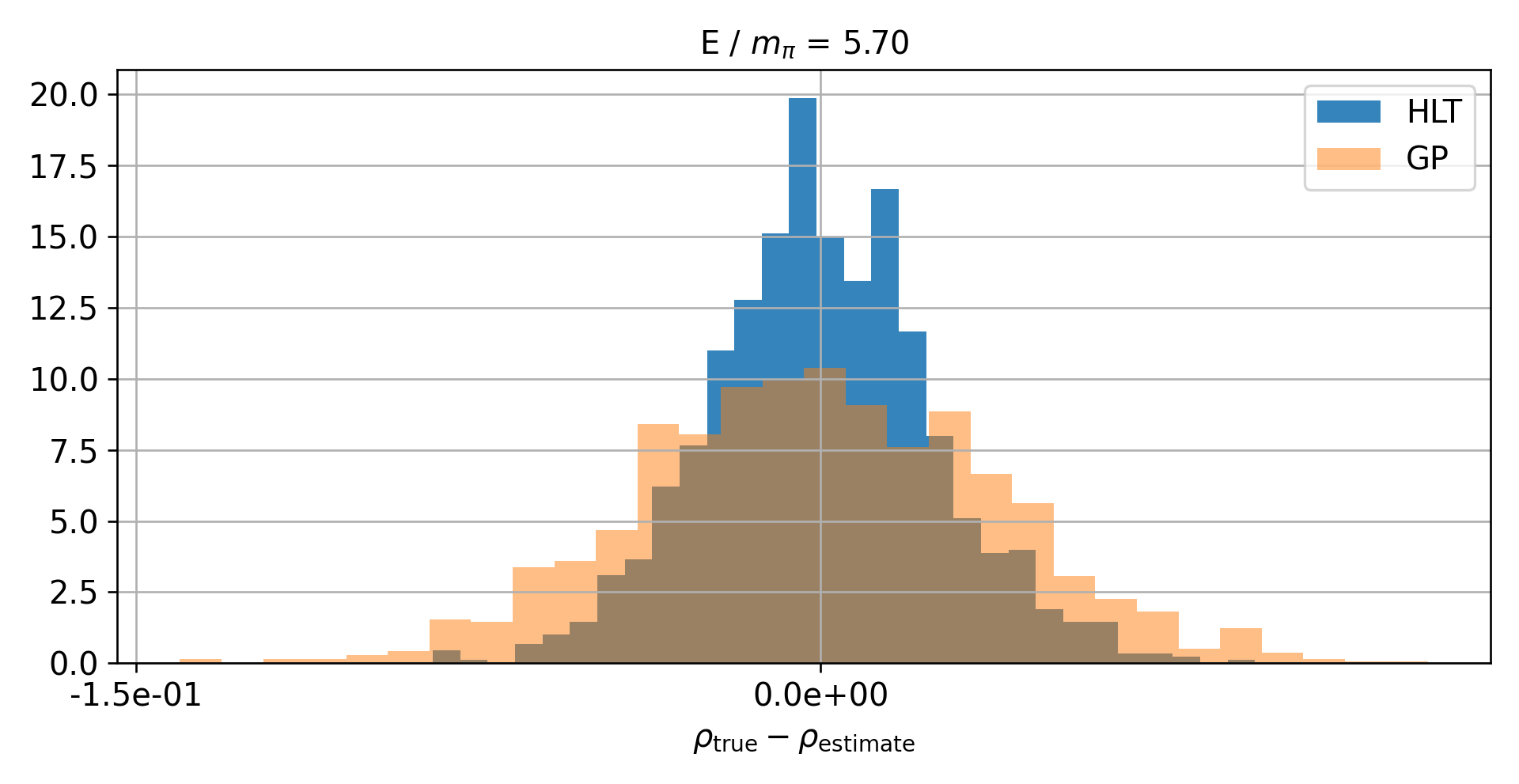}
    \caption{Probability densities for the pull variable defined in Eq.~\eqref{eq:pull} on the left, and the difference between the prediction and the true result on the right. These histograms result from $\simeq 1500$ instances of the inverse problem.}
    \label{fig:histograms}
\end{figure}

We have solved the inverse problem $N_{\rm toys} \simeq 1500$ times with the HLT and with the Bayesian method, using what we referred to as explicit smearing.  In both cases, the prediction was for the spectral density of a vector-vector correlation function of mesons built with light quarks, smeared with a Gaussian with smearing radius $\sigma \simeq 2 m_\pi$. In showing our results, we first assess our ability to remove the bias introduced when regularising the inverse problem. Fig.~\ref{fig:lambda_scan_instances} shows some instances randomly extracted from our validating procedure. The figure shows, in each panel, the stability analysis using three different values of $\alpha$ (cf. Eqs.~\eqref{eq:regulated_coefficients} and~\eqref{eq:the_covariance_is_a_delta}). The panels also report the correct value (horizontal black line) and the predictions from the minimisation of the NLL (Bayes' band) and due to the plateau analysis (HLT band).

The result over the full statistics of over a thousand instances of the inverse problem can be seen in Fig.~\ref{fig:histograms}, where we display on the left panel the probability density of the pull variable,
\begin{equation}\label{eq:pull}
    p_\sigma(E) = \frac{\rho^{\rm pred}_\sigma(E) - \rho^{\rm true}_\sigma(E)}{\Delta \rho_\sigma} \, ,
\end{equation}
where the quantity at the denominator represents the full error on the smeared spectral density. On the right, the histograms of the difference $\rho^{\rm pred}_\sigma(E) - \rho^{\rm true}_\sigma(E)$ is also reported. Differences between histograms obtained with the HLT and the Bayesian analysis can arise for only two reasons. Firstly, the error is computed in different ways. Secondly, the procedure for selecting $\lambda$ is different, leading to different central values. The latter is the only way to justify the difference in the histograms displayed in the right panel of Fig.~\ref{fig:histograms}, since the error does not enter there. The histograms on the right, showing the pull variable, additionally account for the difference in the errors.

The conclusion is that despite at fixed $\lambda$ the Bayesian estimate is more conservative, when coupled to the selection process for the value of $\lambda$, the stability analysis statistically favours smaller values, leading to larger statistical errors and therefore more conservative results. Additionally, as shown in the probability density of the difference (right panel of the same figure), the central value obtained with the stability analysis is statistically closer to the correct one. The same features can also be appreciated in Fig.~\ref{fig:lambda_scan_instances}. It is understood that these results are specific to our data, and in a different context the situation may vary. Nonetheless, the strategy shown here can be used in any case to validate a chosen procedure. More details can be found in Ref.~\cite{DelDebbio:2024lwm}.

\section{Conclusion}

In this proceeding, we reviewed two popular frameworks for the computation of smeared spectral densities from lattice correlators: the Backus-Gilbert method introduced in Ref.~\cite{Hansen:2019idp}, and variations of Bayesian inference using Gaussian Processes~\cite{DelDebbio:2024lwm}. We have shown that when the latter is engineered to compute the probability density for a spectral function smeared with a chosen kernel, the results are intimately related to those of the HLT method. There are two differences. First, the error is computed differently, usually from a resampling procedure in any Backus-Gilbert approach, while it is analytically given from the covariance of the posterior in the Bayesian case. Additionally, the input parameters are selected differently: from a stability analysis in the HLT, from a likelihood function otherwise. Tests on lattice data performed in Ref.~\cite{DelDebbio:2024lwm} suggest that the approaches are loosely compatible.

We then described a way to systematically validate a certain algorithm given a set of lattice correlators for which the covariance matrix has been computed. The strategy is not new, since it is inspired by the so-called ``closure tests'' performed in the contest of the determination of parton distribution functions~\cite{DelDebbio:2021whr}. By using the covariance matrix measured in a recent lattice calculation~\cite{ExtendedTwistedMassCollaborationETMC:2022sta}, we have found that the stability analysis provides on average results that are closer to the correct ones. It is understood that the situation may vary when different data is analysed.

A direction for improvements could be to study the effect of errors on the covariance matrix used in the validation strategy. It is well known that covariance matrices can be hard to estimate reliably, and the impact this may have on our procedure is not known. Additionally, one could improve the Bayesian analysis by sampling the likelihood instead of selecting its maximum value. We leave these ideas for future studies.

While in this work we described two approaches that are based on very different philosophies, they are both linear solutions to the inverse problem. An outlook to non-linear methods can be obtained by looking at machine-learning techniques, such as the one developed in Ref.~\cite{Buzzicotti:2023qdv}. The authors found the results to be in agreement with those obtained with the HLT method.\\

The consistency observed across different approaches is encouraging.

\section{Acknowledgments}

A.L. is funded in part by l'Agence Nationale de la Recherche (ANR), under grant ANR-22-CE31-0011. L.D.D. is funded by the UK Science and Technology Facility Council (STFC) grant ST/P000630/1 and by the ExaTEPP project EP/X01696X/1. M.P. has been partially supported by the Italian PRIN ``Progetti di Ricerca di Rilevante Interesse Nazionale -- Bando 2022'', prot. 2022TJFCYB, by the Spoke 1 ``FutureHPC \& BigData'' of the Italian Research Centre in High-Performance Computing, Big Data and Quantum Computing (ICSC), funded by the European Union -- NextGenerationEU, and by the SFT Scientific Initiative of the Italian Nuclear Physics Institute (INFN). N.T. is supported by the Italian Ministry of University and Research (MUR) under the grant PNRR-M4C2-I1.1-PRIN 2022-PE2 ``Non-perturbative aspects of fundamental interactions, in the Standard Model and beyond'', F53D23001480006, funded by E.U. -- NextGenerationEU.

\bibliographystyle{JHEP}
\bibliography{main.bib}

\end{document}

%% file: shortcuts.tex
\renewcommand{\vec}{\boldsymbol}

\newcommand{\smatrix}{\begin{pmatrix}}
\newcommand{\cmatrix}{\end{pmatrix}}

\DeclareMathOperator*{\argmin}{arg\,min}
